\documentclass[dvips,twocolumn,showpacs,aps,pra,eqsecnum,floatfix]{revtex4} %
\usepackage{graphicx}
\usepackage{pstricks}
\usepackage{amsmath,amsfonts,amsthm,eucal,amssymb}
\begin{document}


\title{\bf A study of the bound states for square potential wells with
position-dependent mass}

\author{A. Ganguly\footnote{On leave of absence from
 City College (C.C.C.B.A), University of Calcutta, 13 Surya Sen Street, Kolkata--700012,
India. {\bf gangulyasish@rediffmail.com}}, \c{S}. Kuru\footnote{On
leave of absence from Department of Physics, Faculty of Science,
Ankara University 06100 Ankara, Turkey. {\bf
kuru@science.ankara.edu.tr}}, J. Negro\footnote{\bf
jnegro@fta.uva.es}, L. M. Nieto\footnote{\bf
luismi@metodos.fam.cie.uva.es}}

\affiliation{Departamento de F\'{\i}sica Te\'orica, At\'omica y
\'Optica, Universidad de Valladolid, 47071 Valladolid, Spain}

\begin{abstract}
 A square potential well with position-dependent mass is studied for bound
 states. Applying appropriate matching conditions, a
 transcendental equation is derived for the energy eigenvalues. Numerical
 results are presented graphically and the variation of the
 energy of the bound states are calculated
 as a function of the well-width and mass.
\end{abstract}



To be published in Phys. Lett. A
 \maketitle

\section{Introduction}
\label{intro}

The concept of position-dependent mass comes from effec\-tive-mass
approximation of many-body problem in condensed matter physics
\cite{wan,sla,ben,bas,von,mor,ein,lev}. In recent times a good number of articles have been
published \cite{dut,roy,koc,gon,alh,bag,q1,q2,mus,san} in this field. The Schr\"odinger equation with
position-dependent mass has been studied in the contexts of
supersymmetry, shape-invariance, Lie algebra, point-canonical
transformation, etc. It is well known that the kinetic energy
operator in this case belongs to the two-parameter family \cite{von}
 \begin{equation}\label{t}
 T(x)=\frac{1}{4} (m^\alpha \,p \, m^\beta\,p \, m^\gamma+m^\gamma \,p \,
 m^\beta \,p \, m^\alpha),
 \end{equation}
where $m=m(x)$ and $p=-i\hbar\, d/dx$, with the constraint
\begin{equation}\label{con2}
  \alpha+\beta+\gamma=-1.
\end{equation}
However, the correct values of the parameters $\alpha,\beta,\gamma$
for a specific model is a long-standing debate
\cite{ben,bas,von,mor,ein,lev}. For example, in the case of a
potential step and barrier \cite{lev}, and in some one-dimensional
potential wells \cite{plas} with varying mass, the following kinetic
energy operator was chosen
\begin{equation}\label{kinp}
  T(x)=\frac{1}{2}\left( p\,\frac{1}{m} \, p \right),
\end{equation}
which corresponds to $\alpha=\gamma=0$, $\beta=-1$. Then, it
was shown that the reflection and transmission
coefficients \cite{lev}, as well as the discrete spectrum \cite{plas}, are
different compared to the constant-mass problem.

In this article we are going to study the following physical
problem: the potential energy has the form of a well, both in
symmetric as well as asymmetric form, and the kinetic energy is
given by (\ref{t}) with a mass function $m(x)$ which has different
constant values inside and outside the well. This problem has
also interest from the point of view of applications in carbon
nanotubes and quantum dots, as can be seen in \cite{biw,yak}. Our
purpose is first to find an ordering of the kinetic energy term
appropriate to this problem, and then, to study in full detail the bound
states of this model and to compare the results with the
conventional constant-mass case.

Several problems which resemble our model have been considered
previously in the literature. For example, the scattering states of
a potential step or barrier were studied in \cite{lev,koc2}.
Other special square well potentials  were analysed in \cite{ein2},
and the bound states for some finite and infinite wells were studied
using other kinetic energy operator and matching conditions in
\cite{plas}. However, our approach is different from the very
beginning because we propose an ordering for kinetic term based on a specific argument
for the matching conditions. This ordering is also used in
\cite{zhu} for studying connection rules for effective-mass wave
functions across an abrupt heterojunction.

The structure of this article is as follows.
Sec.~\ref{Matching} is devoted to fix the ordering in the kinetic term.
In Sec.~\ref{squarewell} a transcendental
equation determining the energy values for the bound states in the
case of a potential energy well is derived. The numerical results are shown
graphically and discussed in Sec.~\ref{Numerical}. Finally,
Section~\ref{Conclusions}  contains the conclusions of our work.

\section{Matching conditions at the discontinuities of the mass and potential function}
\label{Matching}

The Hamiltonian operator for the position-dependent mass problem is
given by
 \begin{equation}\label{h}
 H=T(x)+V(x)
 \end{equation}
where  $T(x)$ is the operator defined in (\ref{t}) and $V(x)$ denotes the
potential term. The one-dimensional
 time-independent Schr\"{o}dinger equation for the stationary states is
 \begin{equation}\label{s}
 H\,\psi(x)=E\,\psi(x).
 \end{equation}
 Substituting (\ref{t}) and (\ref{h}) into (\ref{s}) and taking units such that
 $\hbar^2=2$, we get the
 Schr\"{o}dinger equation for the generalized kinetic energy operator
 \begin{equation}\label{s1}
\frac{d^2\psi}{dx^2}-\frac {m'}{m} \frac{d\psi}{dx} +\left[\frac{1}{2}
\left( \nu\frac {m''}{m}-\eta\,\frac{m'^2}{m^2} \right)+ m(E-V)\right] \psi=0
 \end{equation}
where
\begin{equation}\label{en}
\eta=\alpha(\gamma+2)-\gamma(\alpha+2),\quad\nu=\alpha+\gamma,
\end{equation}
and $m'=dm/dx$. If in Eq. (\ref{s1}) we make the following
transformation
\begin{equation}\label{fi}
\psi(x)=m(x)^{1/2}\phi(x)
\end{equation}
we can eliminate the first order derivative of $\psi$ with respect
to $x$, to arrive at
\begin{equation}\label{s2}
\frac{d^2\phi}{dx^2}+\left[(1+\nu)\frac
{m''}{2m}-\left(\frac{3}{4}+\frac{\eta}{2}\right)\frac
{m'\,^2}{m^2}+m\,(E-V)\right]\phi=0.
\end{equation}

Our aim is to generalize the usual matching conditions for the wave
function $\psi(x)$ suitable for this position-dependent mass
problem. We will focus on the case of mass discontinuities.

\subsection{Matching conditions}
\label{subMatching}

Now, let us assume that the mass $m(x)$ has a finite discontinuity
at $x=a$  of the form
\begin{equation}\label{mt}
m(x)=m_1(x) \Theta(-x+a)+m_2(x)\Theta(x-a)
\end{equation}
where $m_1(x)$ and $m_2(x)$ are smooth functions and the unit step
function $\Theta(x)$ is defined as
\begin{equation}\label{te}
\Theta(x)= \left\{
\begin{array}{ll}
1,\ & x>0\\ [0.5ex]
0,\ & x<0.
\end{array}
\right.
\end{equation}
If we use (\ref{mt}) in Eq. (\ref{s2}), we see that it leads to
strong discontinuities at $x=a$ as well as some terms that require a
careful interpretation. For instance, we should take into account that
$\Theta'\,(x)\equiv d \Theta(x)/d x=\delta(x)$, where $\delta(x)$
denotes the Dirac delta distribution. In order to eliminate these
problems in Eq. (\ref{s2}), we can choose
\begin{equation}\label{con}
1+\nu=0, \quad \frac{3}{4}+\frac{\eta}{2}=0,
\end{equation}
where $\eta$  and $\nu$ were given in (\ref{en}). From these
conditions, we obtain the values of the parameters
$\alpha=\gamma=-\frac{1}{2}$ and $\beta=0$. With such values, the
kinetic energy operator (\ref{t}) becomes
\begin{equation}\label{tn}
T=\frac{1}{2} \left( \frac{1}{\sqrt{m}}\,p^2\frac{1}{\sqrt{m}} \right)
\end{equation}
and Eq. (\ref{s2}) takes the following simple form
\begin{equation}\label{s3}
-\frac{d^2\phi}{dx^2}+
m\,(V-E)\phi=0.
\end{equation}
Now, it is easy to get the matching conditions for $\psi$ at $x=a$.
To do this, first we integrate Eq. (\ref{s3}) around the
discontinuity point $x=a$
\begin{equation}\label{cont1}
\phi'(a+h) - \phi'(a-h)= \int^{a+h}_{a-h}m(x)\,(V(x)-E)\phi(x)\,dx.
\end{equation}
In the interval $(a-h,a+h)$, the functions $m(x)$ and $V(x)$ have
finite discontinuities at $x=a$ and $\phi(x)$ is bounded.
Therefore, when $h$ goes to zero, the integral at the r.h.s. of
(\ref{cont1}) tends to zero. This means that $\phi'(x)$ (and also
$\phi(x)$) will be continuous at $x=a$. In conclusion, using the
transformation (\ref{fi}) we have arrived at the following
matching conditions:  the wave function $\psi(x)$ is such that the
two functions $\psi(x)/\sqrt{m(x)}$ and $(\psi(x)/\sqrt{m(x)})'$
are continuous at $x=a$. Mathematically it may be expressed as
\begin{equation*}
\left .\frac{\psi(x)}{\sqrt{m(x)}}\right |_{x=a-0}=\left
.\frac{\psi(x)}{\sqrt{m(x)}}\right
|_{x=a+0}=\frac{\psi(a)}{\sqrt{m(a)}}\, ,
\end{equation*}

\vspace*{-.5cm}
 \noindent
\begin{equation}\label{cont2}
\left .\left( \frac{\psi(x)}{\sqrt{m(x)}} \right)'\right
|_{x=a-0}\hspace{-.5cm}=\left .\left( \frac{\psi(x)}{\sqrt{m(x)}}
\right)'\right |_{x=a+0}\hspace{-.5cm}=\left .\left(
\frac{\psi(x)}{\sqrt{m(x)}} \right)'\right |_{x=a}\, ,
\end{equation}
where $a$ is an interior point in the domain of the problem. It is
straightforward to check that the above conditions are consistent
with the definition of a time-independent inner product in the
space of square integrable eigenfunctions $\langle
\psi(x),\phi(x)\rangle=\int\psi(x)^*\phi(x)\,dx$,   as well as
with the conservation of the current density
\begin{equation}\label{cu}
j(x)=-i\left(\psi(x)^*\frac{1}{m(x)}\frac{d\psi(x)}{dx}-\frac{d\psi(x)^*}{dx}\frac{1}{m(x)}\psi(x)\right).
\end{equation}
It is clear that these matching conditions are consistent with the Hermitian character of the Hamiltonian (\ref{h}) with
 (\ref{t}). We should stress that the above mentioned boundary conditions indeed define a self-adjoint
 problem (this issue will be addressed in detail elsewhere \cite{gadella}).

\section{A square potential well and a step mass}
\label{squarewell}
\subsection{Asymmetric well}
\label{subAsymmetric}

Let us consider an asymmetric well of the form
\begin{equation}\label{v}
V(x)= \left\{
\begin{array}{ll}
V_1,\ &\ x<-a\\[0.5ex]
0,\  &|x|<a\\[0.5ex]
V_2,\ &\ a<x
\end{array}
\right.
\end{equation}
with the position-dependent mass
\begin{equation}\label{m}
m(x)= \left\{
\begin{array}{ll}
m_1,\ &|x|>a\\[0.5ex]
m_2,\  &|x|<a
\end{array}
\right.
\end{equation}
where $m_1$, $m_2$, $V_1$, and $V_2$ are constants such that $V_2\geq
V_1$, $m_1, m_2>0$, and $m_1\neq m_2$. Next, we will study the bound
states  ($0<E<V_1\leq V_2$) for this problem using to the matching
condition obtained in  the previous section.

The Schr\"{o}dinger equation (\ref{s1}) for the wavefunction
$\psi(x)$ has the following form in each region
\begin{eqnarray}
\label{eq1} \hspace{-.5cm}\frac{d^2\psi}{dx^2}-k^2_1\psi=0, &
k_1=\sqrt{m_1(V_1-E)} , & \  x<-a,
\\
\label{eq2} \hspace{-.5cm}\frac{d^2\psi}{dx^2}+k^2_2\psi=0, &
k_2=\sqrt{m_2\,E} , \hskip1.1cm &  |x|<a,
\\
\label{eq3} \hspace{-.5cm}\frac{d^2\psi}{dx^2}-k^2_3\psi=0, &
k_3=\sqrt{m_1(V_2-E)} , & \  x>a.
\end{eqnarray}
The physical solutions of these equations take the form
\begin{equation}\label{gs}
\psi(x)= \left\{
\begin{array}{ll}
A\,e^{k_1x}, &\ x<-a\\[0.5ex]
C\,\sin(k_2x+\theta),\  &|x|<a\\[0.5ex]
B\,e^{-k_3x},  &\ x>a
\end{array}
\right.
\end{equation}
where $A,B,C$, and $\theta$ are constants to be determined by the
boundary and matching conditions. Using the continuity conditions
(\ref{cont2}) for this solution at $x=-a$, we obtain the following
two equations
\begin{eqnarray}
\label{ac1} A\left(\frac{m_2}{m_1}\right)^{1/2}e^{-k_1a}
&=&C\sin(-k_2a+\theta)
\\
\label{ac2}
A\left(\frac{m_2}{m_1}\right)^{1/2}k_1e^{-k_1a}&=&Ck_2\cos(-k_2a+\theta)
\end{eqnarray}
and at $x=a$, we get
\begin{eqnarray}\label{bc1}
B\left(\frac{m_2}{m_1}\right)^{1/2}e^{-k_3a}&=&C\sin(k_2a+\theta)
\\
\label{bc2}
B\left(\frac{m_2}{m_1}\right)^{1/2}k_3e^{-k_3a}&=&Ck_2\cos(k_2a+\theta).
\end{eqnarray}
From these four equations, we find
\begin{equation}\label{k12}
k_1=k_2\cot(-k_2a+\theta),\quad
k_3=-k_2\cot(k_2a+\theta)
\end{equation}
or
\begin{eqnarray}\label{k12s}
\sin(-k_2a+\theta)&=&\frac{k_2}{\sqrt{m_1V_1-k^2_2(\frac{m_1}{m_2}-1)}}
\\
\label{k13s}
\sin(k_2a+\theta)&=&-\frac{k_2}{\sqrt{m_1V_2-k^2_2(\frac{m_1}{m_2}-1)}}.
\end{eqnarray}
Now, if we eliminate $\theta$ from (\ref{k12s}) and (\ref{k13s}), we get the
transcendental equation
\begin{eqnarray}
2k_2a&=& n\pi-\sin^{-1}\frac{k_2}{\sqrt{m_1V_2-k^2_2(\frac{m_1}{m_2}-1)}}
\nonumber \\
&& -
\sin^{-1}\frac{k_2}{\sqrt{m_1V_1-k^2_2(\frac{m_1}{m_2}-1)}},
\label{e1}
\end{eqnarray}
where inside the well, the momentum $k_2$ of the bound states is
obtained from the values $n=1,2,3,\dots$, and the range of the
inverse sine is taken between $0$ and $\frac{\pi}{2}$. Since
$E=(k_2)^2/m_2$, the roots of this equation also give  the energy
values of the bound states.

As the maximum value that $k_2$ could take is $\sqrt{m_2\,V_1}$,
from Eq. (\ref{e1}), we get an inequality for the number of
bound states
\begin{equation}\label{nb}
2a\sqrt{m_2\,V_1}>\left(n-\frac{1}{2}\right)\pi-\sin^{-1}\frac{\sqrt{m_2\,V_1}}
{\sqrt{m_2V_1+m_1 \Delta V}},
\end{equation}
with $\Delta V=V_2-V_1\geq0$. This means that the total number of
bound states $N$ will be the highest $n$ satisfying this inequality.
When we consider (\ref{nb}) as an equation for $N=1,2,\dots$,
\begin{equation}\label{Nb}
2a\sqrt{m_2\,V_1}=\left(N-\frac{1}{2}\right)\pi-\sin^{-1}\frac{\sqrt{m_2\,V_1}}
{\sqrt{m_2V_1+m_1 \Delta V}}
\end{equation}
we say that the parameters of the well are``critical", in the sense
that by modifying slightly their values we will have one bound state
less or one bound state more.

From (\ref{Nb}), for $N=1$ we see that the first bound state of the
asymmetric square well will appear when the following condition is
satisfied:
\begin{equation}\label{nbb}
\sin^{-1}\frac{\sqrt{m_2\,V_1}}
{\sqrt{m_2V_1+m_1 \Delta V}}=\frac{\pi}{2}-2a\sqrt{m_2\,V_1}\ .
\end{equation}
Therefore, we can say that for  $V_1\neq V_2$, there are always some
values of the parameters of the well that do not allow for bound
states, as it is also the case for the conventional constant-mass
problem.

The influence of $m_1$ (the mass outside the well) on the number of
bound states is not so important. Let us call $f(m_1)$ the function
on the l.h.s. of (\ref{nbb}), then
\begin{equation}\label{nbbb}
\lim_{m_1\to 0} f(m_1) - \lim_{m_1\to \infty} f(m_1)=\frac{\pi}{2}
\ .
\end{equation}
This means that as we increase $m_1$ the spectrum of bound states
either will remain the same or will have one value less. This is in
sharp contradistinction to the influence of $m_2$, that can change any number
of levels in the discrete spectrum.

The conventional constant-mass case is obtained by taking
$m_1=m_2=m$ in Eq. (\ref{e1}) to get the well known formula
\cite{landau}
\begin{equation}\label{e2}
2k_2a=n\pi-\sin^{-1}\frac{k_2}{\sqrt{mV_2}}-\sin^{-1}\frac{k_2}{\sqrt{mV_1}},
\end{equation}
where $n=1,2,3,\dots$ The inequality for the number of bound states
is obtained now from (\ref{nb}) if we put $m_1=m_2=m$:
\begin{equation}\label{nbm}
2a\sqrt{m\,V_1}>\left(n-\frac{1}{2}\right)\pi-\sin^{-1}\sqrt{\frac{V_1}{V_2}}.
\end{equation}
However, we must stress that in the position-dependent mass case the
critical values given by formula (\ref{nb}) for the number of bound
states depend on both $m_1$ and $m_2$, in particular, as $m_2\to0$,
we see from (\ref{nbb}) that no bound states will remain in the
well.

\subsection{Symmetric well}
\label{subsymmetric}

If $V_1=V_2=V$, from (\ref{k12}), we get the energy equations
corresponding to even and odd eigenfunctions by replacing
$\theta=\pi/2$ and $\theta=0$, respectively
\begin{equation}\label{ke}
\frac{k_1}{k_2} =\tan{k_2 a}, \qquad \frac{k_2}{k_1} =-\tan{k_2 a}.
\end{equation}
In this case, we can also write for both even and odd solutions
the transcendental Eq. (\ref{e1}), using $k_2=\sqrt{m_2E}$, in the
following way
\begin{equation}\label{e4}
2k_2a = n\pi-2 \sin^{-1}\frac{k_2}{\sqrt{k_2^2+m_1(V-E)}},
\end{equation}
where $n=1,2,3,\dots$ Here, the argument of the inverse sine is
always less than $1$ for $V-E>0$, and therefore the formula is well
defined. If we make $E=V$ in (\ref{e4})  the number of bound states
$N$ is given by the highest $n$ satisfying the inequality
\begin{equation}\label{nbm1}
2a\sqrt{m_2\,V}> (n-1)\pi.
\end{equation}
This means that  there are always bound states for the symmetric
well. Even more, once the depth $V$ and the width $2a$ of the well
are fixed, the number of bound states does not depend on $m_1$, it
depends only on the mass inside of the well ($m_2$). The
relation (\ref{nbm1}) for the number of bound states coincides with
that obtained also by Plastino {\it et al} using different
matching conditions. However, the energy equations
(\ref{ke})--(\ref{e4}) are different from \cite{plas}. In fact, when the inside
mass is bigger (smaller) than the outside mass, then our energy
spectrum is lower (upper) than the spectrum given in \cite{plas}.

The critical values of the well (which determine when we have one
bound state more) are obtained from
\begin{equation}\label{bm1}
2a\sqrt{m_2\,V} =(N-1)\pi.
\end{equation}
Thus the critical values of $a$ are linear in the number $N-1$ of
bound states, while the critical values $m_2$ grow as the square of
$(N-1)$. In conclusion, the important parameter here is the
mass $m_2$ inside the well, while the mass outside ($m_1$) plays a
secondary role.

The number of bound states for the conventional
constant-mass symmetric well are given also by (\ref{nbm1})
replacing $m_2\to m$. However, the energy of these bound states is
defined through (\ref{e4}) by choosing $m_1=m_2=m$
\begin{equation}\label{e5}
k_2a = n\,\frac{\pi}2-\sin^{-1}\frac{k_2}{\sqrt{m V}}.
\end{equation}
Hence the values of the energies will be different from the
position-dependent mass.

\section{Numerical results and discussion}
\label{Numerical}

In this section we will discuss some numerical results and graphics
obtained from the formulas of the previous sections, paying attention
to the consequences of mass variation.

Consider the situation of a particle of mass $m_1$ in a square well
potential, where we have altered the conditions inside the well to
produce a different effective mass $m_2$. We want to study the
effects that this change of mass will produce on the bound states.

\subsection{Features of the asymmetric well}
\label{subFeaturesasymmetric}

The main characteristics of the asymmetric well may be summarized as follows:

\begin{enumerate}

\item
If $V_1<V_2$, it can be seen from formula (\ref{nb}) that the number
of bound states depends linearly on the width $2a$ of the asymmetric
well as shown in Fig.~\ref{figura1},%
\begin{figure}[htp]
\centerline{
\includegraphics{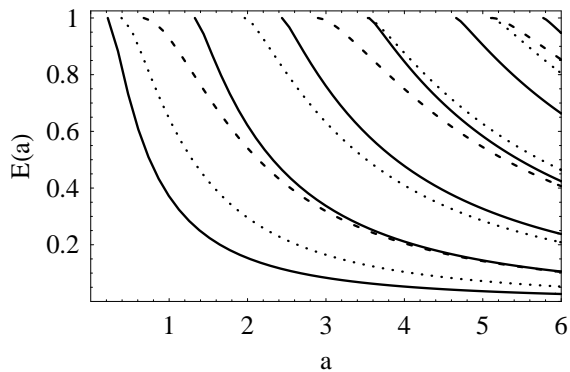}}
\caption{Plot of the energy values and the number of the bound
states depending on the width of the well $2a$ when the other
parameters are fixed as: $V_1=1$, $V_2=2$, $m_1=1$,  $m_2=2$
(solid lines), $m_2=1$ (dotted lines), and  $m_2=0.5$ (dashed
lines).}
  \label{figura1}\end{figure}
 where we have plotted three
situations:

\begin{enumerate}
\item
the usual constant-mass case, $m_1=m_2=1$,
\item
$m_1=1$, $m_2=2$, that is the case where the mass inside is
bigger than outside the well,
\item
$m_1=1$, $m_2=0.5$, that corresponds to a lighter mass inside
the well.
\end{enumerate}

Thus, the effect of $m_2$ on the bound states is quite strong for
any value of $a$, as can be seen in Fig.~\ref{figura1}. When $m_2$
is decreasing, the number of bound states also decrease, while
each energy level is higher. %
\begin{table}[t]
\caption{The critical values of $a$ obtained from Eq. (\ref{Nb}),
determine the width where a new bound state appears in the well.
\label{tabone}}
\begin{ruledtabular}
\begin{tabular}{llllllll}
 $m_1$  & $m_2$ & $a^{(1)}$  & $a^{(2)}$ &  $a^{(3)}$ &  $a^{(4)} $ & $a^{(5)}$ & $a^{(6)}$ \\ \hline
1 & 2 & 0.2176& 1.3283&2.4390&3.5498&4.6605 &5.7712  \\ 
1 & 1 &0.3927&1.9635 &3.5343 &5.1051 &6.6759 &8.2469  \\ 
1 & 0.5 &0.6755 &2.8970 &5.1184 &7.3398 &9.5613 &11.7827  \\ 
\end{tabular}
\end{ruledtabular}
\end{table}
In  Table~\ref{tabone}  it is shown the effect of these three
values of $m_2$ on the first critical values of the width $2a$.

\item
In Fig.~\ref{figura2} %
\begin{figure}[htp]
\centerline{
\includegraphics{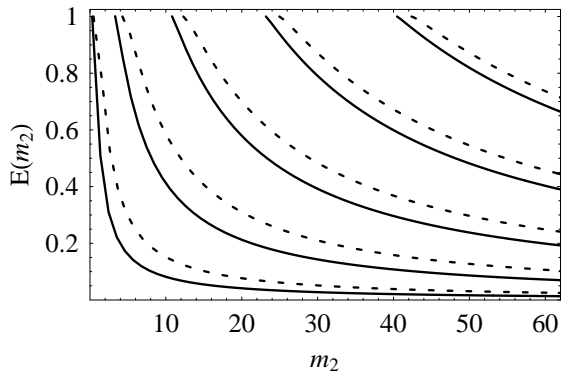}}
\caption{Plot of the energy values and the number of the bound
states depending on the mass $m_2$ when the other parameters are
fixed as: $a=1$, $V_1=1$, $V_2=2$, $m_1=1$ (solid lines), and
$m_1=10$ (dashed lines).}
  \label{figura2}
\end{figure}
 we plotted the bound state energies as we
change only the inside mass $m_2$ for $m_1=1$ and $m_1=10$
(leaving the other parameter fixed). We see that the number of
bound states is almost proportional to the square root of $m_2$
due to (\ref{nb}), while the energy value of each bound state
decreases with $m_2$. The influence of $V_1$ is similar as that of
$m_2$.

\item
In the same figure, we can see that the influence of $m_1$ (the
mass outside the well) on the number of bound states is not so
important. Table~\ref{tabtwo} %
\begin{table}[t]
\caption{The critical values of $m_2$ obtained from Eq.
(\ref{Nb}), determine the mass where a new bound state appears in
the well. \label{tabtwo}} 
\begin{ruledtabular}
\begin{tabular}{lllllll}
 $m_1$  &  $m_2^{(1)}$  &  $m_2^{(2)}$ & $m_2^{(3)}$ & $m_2^{(4)} $ & $m_2^{(5)}$ &  $m_2^{(6)}$\\ \hline
1 &0.2899&3.3189&10.8186 &23.1821 &40.4642 &62.6758  \\
10 &0.4618&4.2715&12.3087&24.9461&42.3710&64.6628 \\
\end{tabular}
\end{ruledtabular}
\end{table}
 shows the influence of $m_1$ on the
first critical energy values of $m_2$. The parameter $V_2$ has
similar qualitative effects on the energy values and the number of
the bound states as the parameter $m_1$.
\end{enumerate}

\subsection{Features of the symmetric well}
\label{subFeaturessymmetric}

In the symmetric well the importance of the mass $m_2$ inside
the well is even stronger than in the asymmetric well. The main characteristics of the symmetric well are the following:

\begin{enumerate}
\item
The number of bound states depends linearly on $a$, the square root
of the $m_2$ and $V$, as in the asymmetric well, but it is
independent of $m_1$.

\item
The dependence on $m_2$ is shown in Fig.~\ref{figura3}, %
\begin{figure}[htp]
\centerline{
\includegraphics{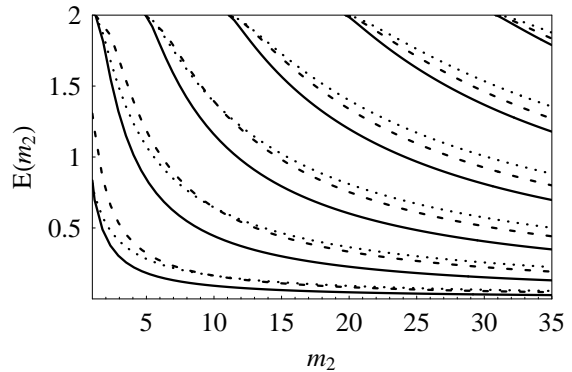}}
\caption{Plot of the energy values and the number of the bound
states depending on the mass $m_2$ when the other parameters are
fixed as: $a=1$, $V_1=V_2=2$, $m_1=1$ (solid lines), $m_1=10$
(dashed lines), and  $m_1=m_2=m$ (dotted lines). The critical
values for the three cases are the same: $m^{(1)}_2=0$,
$m^{(2)}_2=1.23$, $m^{(3)}_2=4.93$, $m^{(4)}_2=11.10$,
$m^{(5)}_2=19.74$, $m^{(6)}_2=30.84$, obtained from Eq.
(\ref{bm1}).} \label{figura3}
\end{figure}
 a detail
is given in Fig.~\ref{figura4}. %
\begin{figure}[htp]
\centerline{
\includegraphics{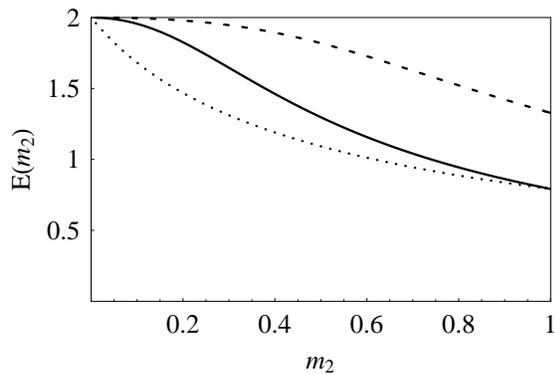}}
\caption{Plot of the energy values and the number of the bound
states depending on the mass $m_2$ when the other parameters are
fixed as: $a=1$, $V_1=V_2=2$, $m_1=1$ (solid lines), $m_1=10$
(dashed lines), and $m_1=m_2=m$ (dotted lines).} \label{figura4}
\end{figure}
 As a consequence, the critical
values of $m_2$ are the same, independently of the values of
$m_1$.

\item
The energy values given by  (\ref{e4}) are only slightly affected
by the variation of $m_1$, the mass outside the well. This is
shown in Fig.~\ref{figura3}, for the values $m_1=1$, $m_1=10$  and
$m_1=m_2=m$. A detail of this plot is shown in Fig.~\ref{figura4},
where the energy levels of the case  $m_1=1$ coincide
with the conventional constant-mass case ($m_1=m_2=m$) at the
point $m_2=1$. {Clearly, the energy levels of the case
$m_1=10$ will also coincide with those for constant-mass
case $m_1=m_2=m$ at the point $m_2=10$ (see Fig.~\ref{figura3}).}

\end{enumerate}

\section{Conclusions}
\label{Conclusions}

In this article we have studied a potential well both in symmetric
and asymmetric form, with different constant mass inside and outside
the well. We have proposed specific values for the ordering parameters in the kinetic
term based on the simple argument that strong singularities in the effective-mass
Schr\"odinger equation should be avoided.
However, this choice is not unique: for instance,
the symmetric well has been considered in \cite{plas} with a kinetic
term proposed in \cite{lev}. Similar to the conventional
constant-mass case we have obtained a transcendental equation for
bound state energies. We have shown our numerical results for
different values of the parameters. Many interesting differences
from conventional constant-mass situation have been pointed out. In
particular, the bound states are controlled mainly by the mass
inside the well, while the mass outside, in general acts simply as a
tuning of the specific values of the energies in the spectrum.
The experiments have also shown that the energy level spacing is
very sensitive to the effective mass inside the well, while the
effects of the other parameters are not so important \cite{biw}.
Finally, we have remarked the influence of the matching conditions,
related to the kinetic term, on the discrete spectrum by comparing
our results with those of \cite{plas}.

We have also determined some critical values of the well, that is the
values of the potential and mass functions giving rise to the new
bound states in the well. These critical values are important in the
study of the scattering states, because they fix the conditions
where the transmission coefficients reach the maximum values.

 \section*{Acknowledgments}
This work has been partially supported by Spanish Ministerio de
Educaci\'on y Ciencia (Projects MTM2005-09183 and FIS2005-03989),
Ministerio de Asuntos Exteriores (AECI grants 0000147625 of
\c{S}.K. and 0000147287 of A.G.), and Junta de Castilla y Le\'on
(Excellence Project VA013C05). A.G. and \c{S}.K. acknowledge the
warm hospitality at Department of Theoretical Physics, University
of Valladolid, Spain, where this work has been carried out. The
authors thank the anonymous referee for his interesting comments
and for pointing out some relevant references.



\begin{thebibliography}{00}

\bibitem{wan}
G.H. Wannier,  {Phys. Rev.}  {\bf 52}, (1937) 191

\bibitem{sla}
J.C. Slater,  {Phys. Rev.}  {\bf 76}, (1949) 1592

 \bibitem{ben}
 D.J. BenDaniel  and C.B. Duke,  {Phys. Rev.} {\bf 152}, (1966) 683

 \bibitem{bas}
 G. Bastard, {Phys. Rev.} B  {\bf 24}, (1981) 5693

 \bibitem{von}
 O. von Roos, {Phys. Rev.} B  {\bf 27}, (1983) 7547

 \bibitem{mor}
 R.A. Morrow, {Phys. Rev.} B  {\bf 35}, (1987) 8074

\bibitem{ein}
G.T. Einevoll  and P.C. Hemmer, {J. Phys. C: Solid State Phys.} {\bf 21}, (1988) L1193

 \bibitem{lev}
 J.-M. L\'evy-Leblond,  {Eur. J. Phys.} {\bf 13}, (1992) 215

\bibitem{plas}
 A.R. Plastino, A. Puente, M. Casas, F. Garcias, and A. Plastino,  {Rev. Mex. Fis.} {\bf 46}, (2000)  78

\bibitem{dut}
A. de Souza Dutra  and C.A.S. Almeida,  {Phys. Lett.} A {\bf 275}, (2000) 25

 \bibitem{roy}
 B. Roy and P. Roy, {J. Phys. A: Math. Gen. } {\bf 35}, (2002) 3961

\bibitem{koc}
R. Ko\c{c}, M. Koca, and E. K\"orc\"uk, {J. Phys. A: Math. Gen.} {\bf 35}, (2002) L527

 \bibitem{gon}
B. Gonul, O. Ozer,  and F. Uzgum,  {Mod. Phys. Lett.} A {\bf 17}, (2002) 2453

 \bibitem{alh}
 A.D. Alhaidari, {Int. J. Theor. Phys.} {\bf 42}, (2003) 2999

 \bibitem{bag}
 B. Bagchi, P. Gorain, C. Quesne, and R. Roychoudhury, Mod. Phys. Lett. A {\bf 19}, (2004) 2765

 \bibitem{q1}
C. Quesne and V.M. Tkachuk,  {J. Phys. A: Math. Gen.} \textbf{37},
(2004) 4267

 \bibitem{q2}
C. Quesne, {Ann. Phys.} (NY) \textbf{321}, (2006) 1221

 \bibitem{mus}
O. Mustafa  and S.H. Mazharimousavi, quant-ph/0603134

 \bibitem{san}
J.F. Cari\~{n}ena, M.F. Ra\~{n}ada, and M. Santander,  {Ann. Phys.} (to be published)

\bibitem{biw}
S.K. Biwas, L.J. Schowalter, Y.J. Jung, and R. Vajtai {Appl. Phys.
Lett.} {\bf 86}, (2005) 183101

\bibitem{yak}
A.I. Yakimov, A.V. Dvurechenskii, A.I. Nikiforov, A.A. Bloshkin,
A.V. Nenashev, and V.A. Volodin { Phys. Rev. B} {\bf 73}, (2006)
115333

\bibitem{koc2}
R. Ko\c{c}, M. Koca, and G. \c{S}ahino\u{g}lu, {Eur. Phys. J. B} {\bf 48}, (2005) 583

\bibitem{ein2}
J. Thomsen, G.T. Einevoll,  and P.C. Hemmer, {Phys. Rev. B} {\bf 39}, (1989) 12783;
G.T. Einevoll, P.C. Hemmer,  and J. Thomsen, {Phys. Rev. B} {\bf 42}, (1990) 3485

\bibitem{zhu}
Q. Zhu, and H. Kroemer, {Phys. Rev. B} {\bf 27}, (1983) 3519

\bibitem{gadella}
M. Gadella, \c{S}. Kuru, and J. Negro, {\it Conditions for self-adjointness of a Hamiltonian with variable mass} (in preparation)

\bibitem{landau}
L.D.  Landau  and  E.M. Lifshitz, {\it Quantum  Mechanics} (Pergamon Press, Oxford, 1965)


\end{thebibliography}
\end{document}